%% file: DPF13-muller.tex
\documentclass[12pt]{article}
\usepackage{graphicx}

\newcommand\pubnumber{DPF2013-127}
\newcommand\pubdate{\today}
\input babarsym

\def\eeqqb  {\ensuremath{\epem \!\!\to \! \qqbar}\xspace}

\def\ppbar  {\ensuremath{p/\overline{p}}\xspace}
\def\plab   {\ensuremath {p_{\rm lab}}\xspace} 
\def\pstar  {\ensuremath {p^*}\xspace} 
\def\thlab  {\ensuremath {\theta_{\rm lab}}\xspace}
\def\ecm    {\ensuremath {E_{\rm CM}}\xspace}

\def\address{SLAC National Accelerator Laboratory\\
Stanford, CA  94309, USA}

\def\Title#1{\begin{center} {\Large #1 } \end{center}}
\def\Author#1{\begin{center}{ \sc #1} \end{center}}
\def\Address#1{\begin{center}{ \it #1} \end{center}}

\newcommand\pubblock{\rightline{\begin{tabular}{l} \pubnumber\\
         \pubdate  \end{tabular}}}
\newenvironment{Abstract}{\begin{quotation}  }{\end{quotation}}
\newenvironment{Presented}{\begin{quotation} \begin{center} 
             PRESENTED AT\end{center}\bigskip 
      \begin{center}\begin{large}}{\end{large}\end{center} \end{quotation}}

\input econfmacros.tex

\begin{document}
\begin{titlepage}
\pubblock

\vfill
\Title{Measurement of Inclusive Production of Light Charged Hadrons at
\babar }
\vfill
\Author{ David Muller\\ For the \babar\ Collaboration}
\Address{\address}
\vfill
\begin{Abstract}
Inclusive hadron production cross sections in \epem collisions shed
light on the fundamental fragmentation and hadronization processes.  
We present measurements of the inclusive spectra of charged pions,
kaons and protons in hadronic events at a center-of-mass energy of
10.54~\gev.
These results are compared with theoretical predictions and the
predictions of three hadronization models.
Along with previous measurements at higher energies,
they are also used to study the scaling properties of hadron
production. 
\end{Abstract}
\vfill
\begin{Presented}
DPF 2013\\
The Meeting of the American Physical Society\\
Division of Particles and Fields\\
Santa Cruz, California, August 13--17, 2013\\
\end{Presented}
\vfill
\end{titlepage}
\def\thefootnote{\fnsymbol{footnote}}
\setcounter{footnote}{0}

\section{Introduction}

The production of ``jets'' of hadrons from energetic quarks and gluons
in high-energy collisions is understood qualitatively, 
but there are few quantitative theoretical predictions.
Detailed measurements of jet structure provide a probe of the
confining property of the strong interaction,
and an empirical understanding of jets is vital to much ongoing work
in high-energy physics, where much known and new physics manifests
iteself in the form of jets.
Identified hadrons probe the dependence of this process on the hadron
mass and quantum numbers, such as spin, baryon number, etc., as well
as on the parton flavor.
Here we present measurements~\cite{babarpikp} of the production of
\pipm, \Kpm and \ppbar in \eeqqb events at a center-of-mass (CM) energy
of $\ecm = 10.54$~\gev, where there are four active quark flavors.
We use a high-quality sample of data recorded with the \babar\ detector, 
which features excellent and well understood tracking and particle
identification.

\section{Analysis}

We select hadronic events with low bias in the track
multiplicity and momentum by making requirements on the event vertex,
topology and visible energy, 
the direction of the thrust axis,
and the \epm content in low-multiplicity events. 
The efficiency is 70\% and the sample of 2.2 million events
contains three nonnegligible backgrounds: 
4.5\% of the events are $\tau$-pairs,
which are well modeled and can be subtracted reliably;
radiative Bhabha events are measured in the data and contribute only
0.1\% of the events, but several percent of the highest-momentum
tracks identified as \pipm;
and two-photon events are limited from the data to a level well below
that of the $\tau$-pairs.

Within these events, 
we select charged tracks with good particle identification information
and low bias on particle type 
by making requirements on the numbers of hits and the
extrapolation to both the event vertex and the particle identification
subsystems. 
The efficiency is about 80\% with a small dependence on momentum \plab
and polar angle \thlab in the laboratory frame,
except for the requirement of 200~\mevc in momentum transverse to the
beam axis.
This efficiency is measured in the data with a relative uncertainty
below 1\% above 1~\gevc, 
and increasing to 2--5\%, depending on particle type, 
as the momentum decreases to 0.2~\gev.
These requirements remove decay products of \KS and weakly decaying
strange baryons, which are sometimes included in such measurements.
We report both ``prompt'' and ``conventional'' results, excluding and
including such tracks, respectively.

Tracks are identified as pions, kaons, or protons using a combination
of energy loss in the tracking chambers and angles measured in the
Cherenkov detector.
The algorithm is optimized for high purity and smooth variation with
\plab and \thlab.
Correct identification efficiencies are above 99\% for \plab below
0.6~\gevc, 
above 90\% below 2.5~\gevc, and then fall off at higher \plab.
Misidentification rates are at most 5\% and typically below 2\%;
not all tracks are identified.

We divide the selected tracks into six regions in \thlab and 
45 bins in \plab.
In each bin and region,
we count the numbers of tracks identified as \pipm, \Kpm and \ppbar, 
apply the inverse of the matrix of (mis)identification efficiencies, 
and check that the sum of the resulting numbers of true \pipm, \Kpm
and \ppbar is consistent with the total number of selected tracks,
within the uncertainties of the efficiency matrix.
Dividing by the number of selected events and the bin width, we obtain
the raw production rates, 
$(1/N_{evt}^{sel}) (dn_j/d\plab)$, $j=\pi, K, p$.

We subtract the backgrounds expected from the three classes of
non-hadronic events discussed above, 
as well as those due to beam-related particles, 
photon conversion and other interactions in the detector material,
and residual decays of \KS and weakly decaying strange baryons.
Dividing by the track finding efficiency and normalizing to the
number of hadronic events in the sample, 
we obtain prompt corrected production rates in the laboratory frame, 
$(1/N_{evt}^{had}) (dn_j/d\plab)$.

In each \thlab region, we transform these rates into the the
\epem CM frame to obtain $(1/N_{evt}^{had}) (dn_j/d\pstar)$.
This includes corrections for resolution, energy loss
and initial state radiation,
and also model-dependent transfer matrices and acceptance factors.
The former are sensistive to the true \pstar distributions, 
which are the objects of this measurement,
so we adopt an iterative procedure in which the simulated \pstar
distribution is reweighted to match the data, 
the matrices are recalculated, and the transformation is repeated.
This procedure converges in two iterations.
The acceptance is insensitive to \pstar but sensitive to the angular
distribution.
We check this by comparing the results from the six \thlab regions, 
which would show a specific pattern of differences if the simulated 
distribution were incorrect. 
The six results are consistent within the expected uncertainties,
and this also limits several other potential systematic effects, 
which would produce different patterns of differences.

\section{Results}

\begin{figure}[htb]
\centering
\includegraphics[width=0.355\hsize]{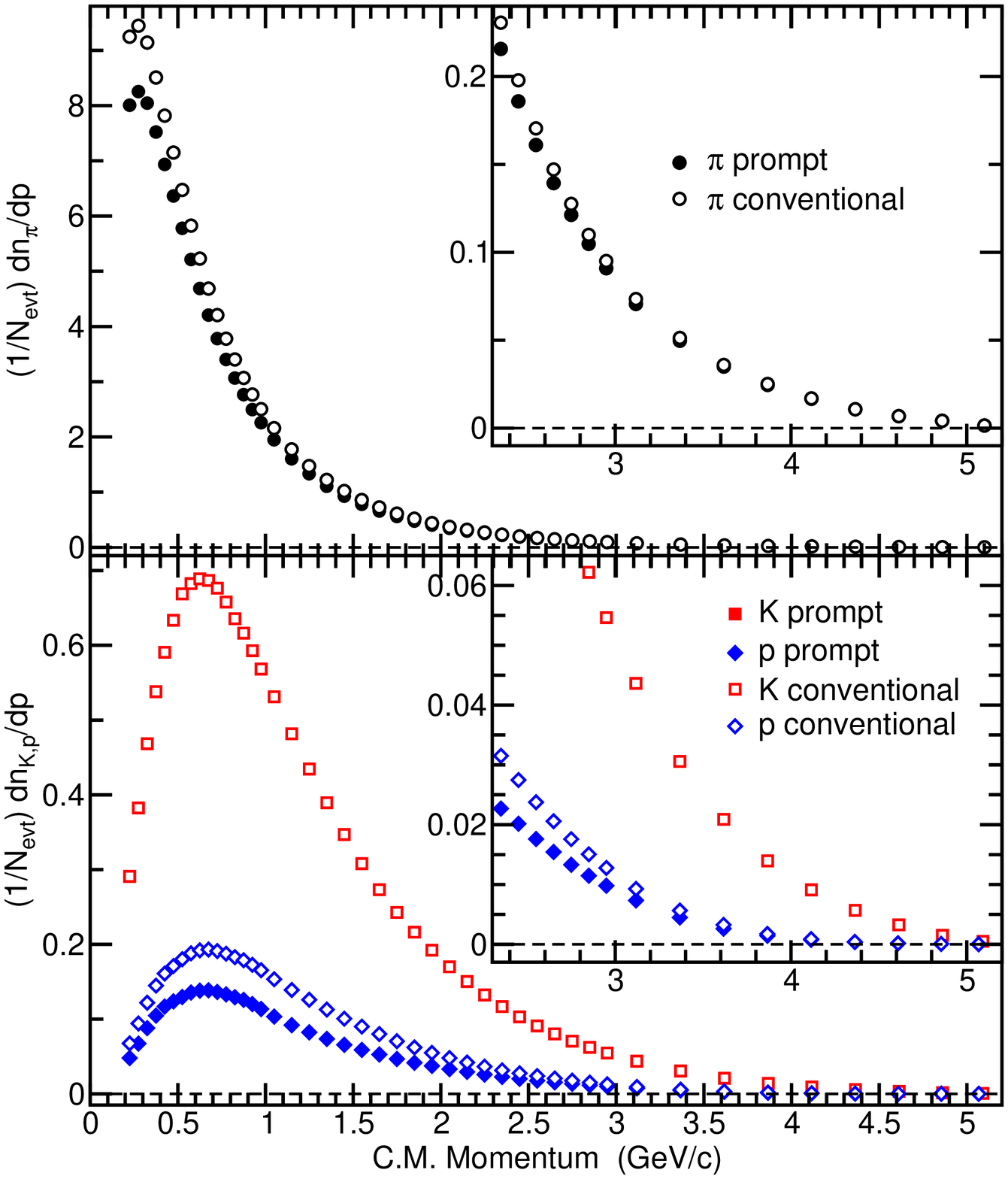}
\includegraphics[width=0.287\hsize]{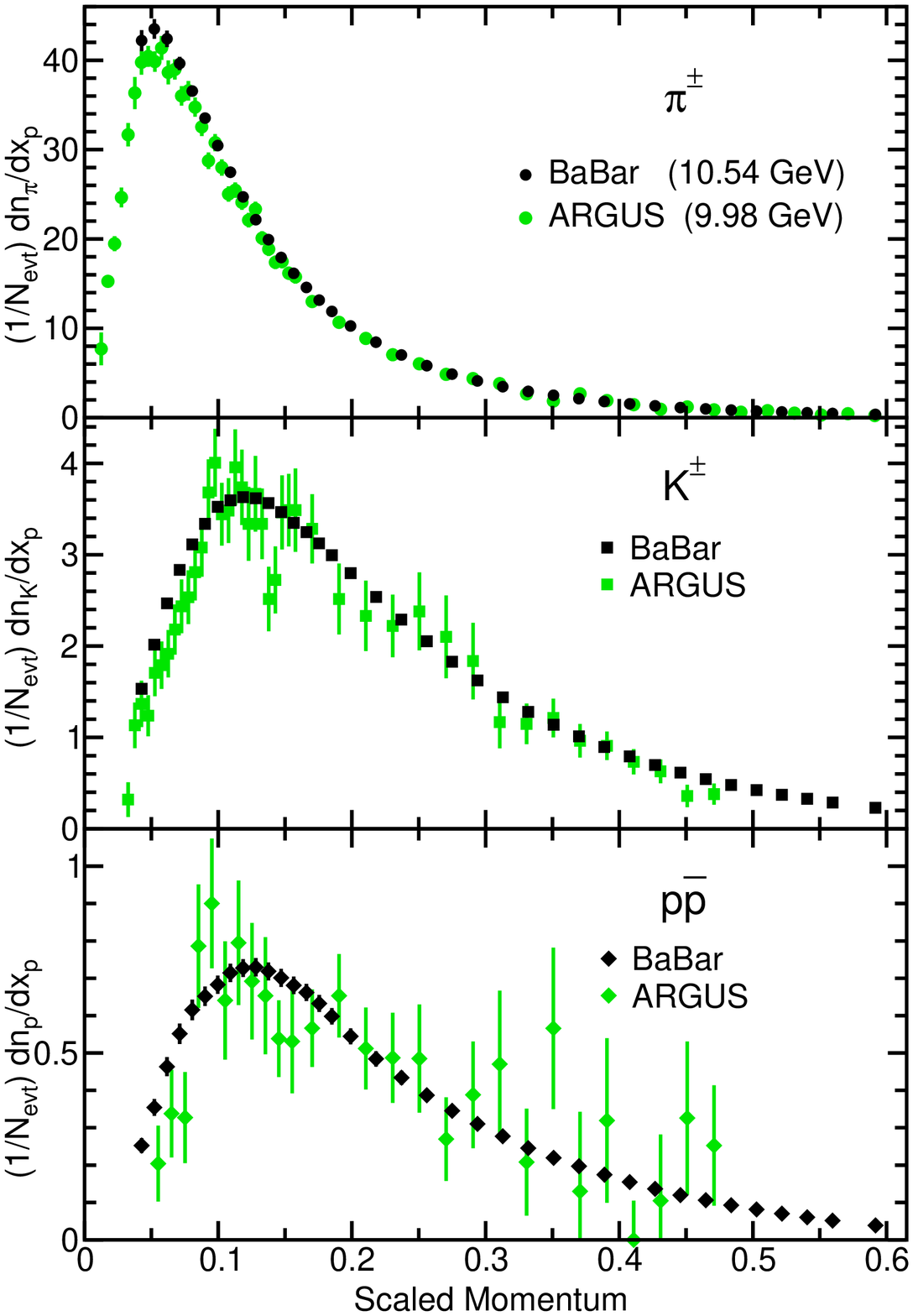}
\includegraphics[width=0.339\hsize]{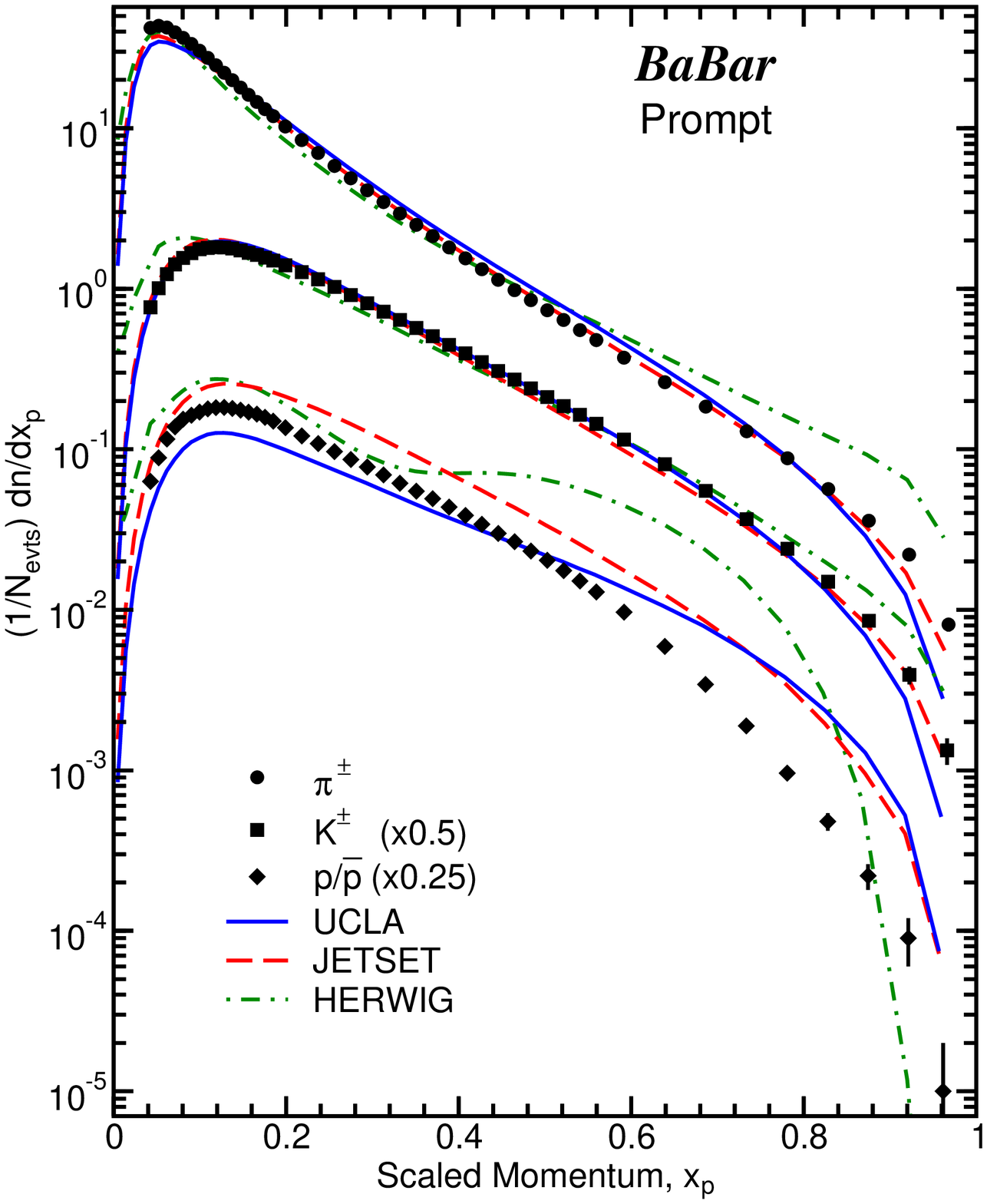}
\caption{
 Left:  Differential production rates per hadronic event per unit
  momentum \pstar in the CM frame for prompt (filled) and conventional
  (open symbols) \pipm (top), \Kpm and \ppbar (bottom).
 Middle:  comparison with results from ARGUS at 9.98~\gev.
 Right:  comparison with hadronization models.}
\label{fig:provcon}
\end{figure}

Averaging over the \thlab regions contributing to each \pstar bin,
we obtain the prompt differential production rates per hadronic event
shown in Fig.~\ref{fig:provcon} as the filled symbols, 
and adding the strange-particle decay products gives the conventional 
rates, shown as the open symbols.
The measurements cover the \pstar range from 0.2~\gevc to the beam
energy, 
including most of the \Kpm and \ppbar spectra and the peak and
high side of the \pipm spectrum.
The prompt and conventional results are indistinguishable for \Kpm, 
as the only difference is from $\Omega^-$ baryon decays.
For \pipm, there is a few-percent difference over most of the range
due to \KS decays, 
and up to 10\% more at low \pstar due to strange baryon decays.
For \ppbar there is a 40\% difference over most of the range.

The statistical uncertainties are smaller than the data points,
and the systematic uncertainties are strongly correlated over both
short and long ranges.
There is a 1\% normalization uncertainty.
Those from tracking are fully correlated,
total as much as 5\% at 0.2~\gevc, 
and decrease rapidly with increasing \pstar.
Those from backgrounds and particle identification grow with
\pstar to as much as 5\% and 50\% in the the highest bin,
and show full and 4--6 bin correlations, respectively.

In Fig.~\ref{fig:provcon} we compare our prompt results with previous
results from the ARGUS experiment~\cite{arguspikp} at the slightly
lower \ecm of 9.98~\gev.  
While our results are much more precise statistically and extend to
higher \pstar,
their low-\pstar \pipm and \Kpm coverage is better, 
and the systematic uncertainties are comparable.
Above 1~\gevc the results are consistent.
Below that, the ARGUS measurements fall below ours as \pstar
decreases, 
consistent with the small scaling violation expected from the 
difference in \ecm.

We compare with the predictions of three hadronization models in
Fig.~\ref{fig:provcon}, using in each case the default parameter values.
Although some models reproduce some spectra at some \pstar values, 
the overall decsription of the data is poor.
Various tuned parameter sets exist and some come closer to the
amplitudes of some spectra, 
but none reproducess the shapes.
Similar differences are seen when comparing conventional spectra,
and these data should prove useful in obtaining better tunes.

\begin{figure}[htb]
\centering
\includegraphics[width=0.325\hsize]{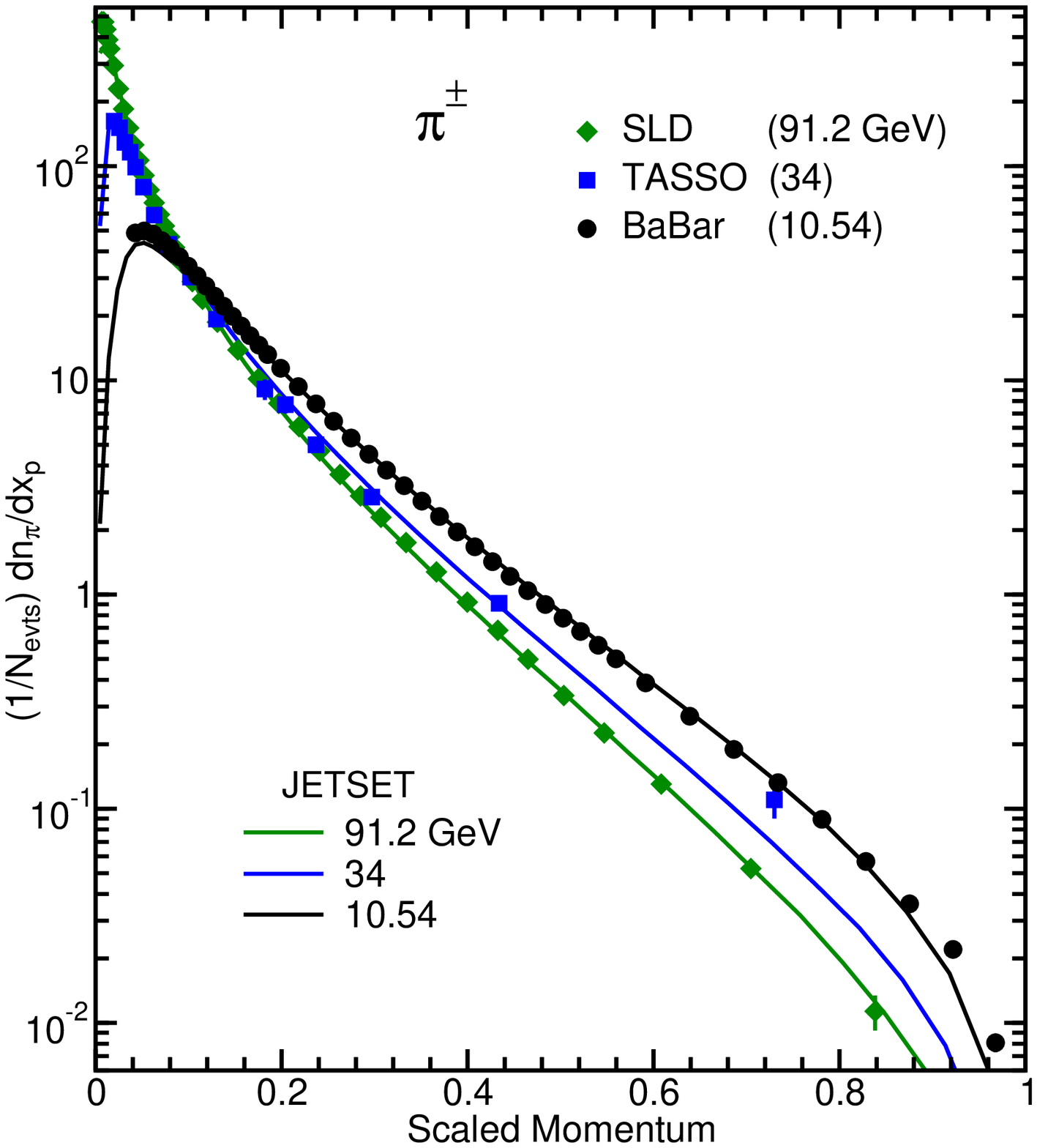}
\includegraphics[width=0.325\hsize]{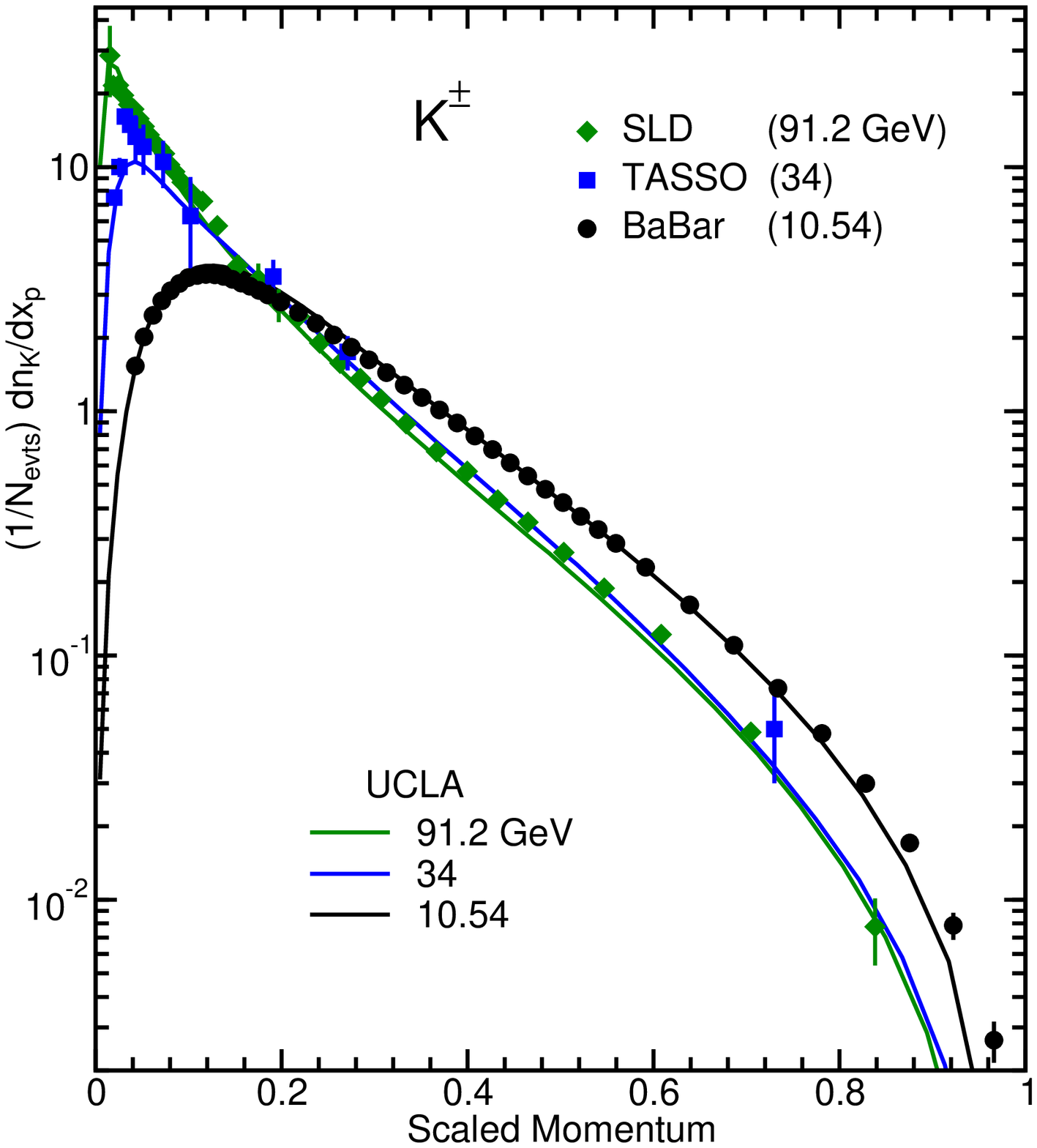}
\includegraphics[width=0.325\hsize]{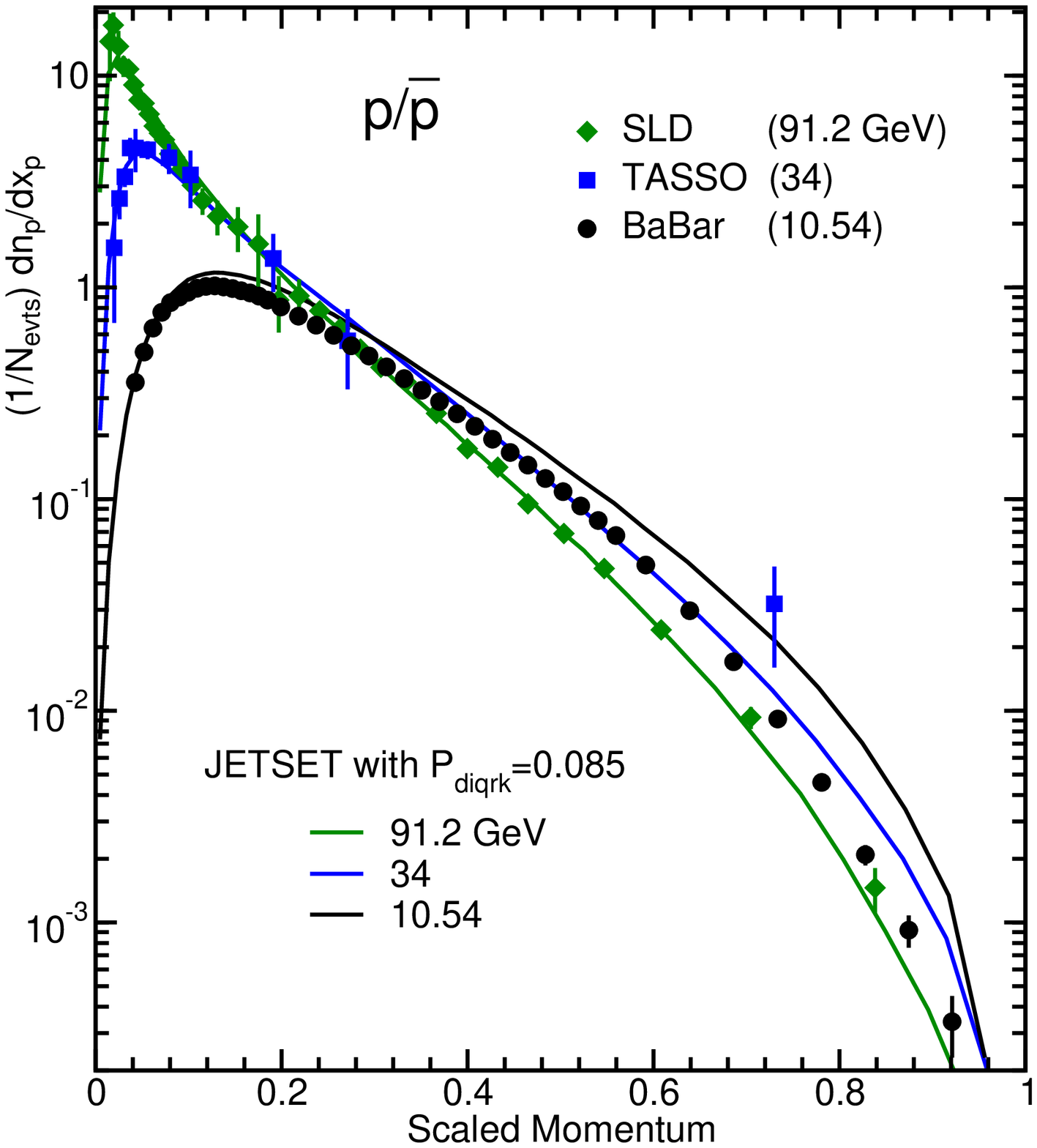}
\caption{
 Conventional \pipm (left), \Kpm (middle) and \ppbar (right) spectra
 measured at three different CM energies, 
 compared with the predictions of selected models.}
\label{fig:scal}
\end{figure}

We study the scaling properties of these spectra by comparing our
conventional results with
data from the TASSO~\cite{tassopikp} and SLD~\cite{sldpikp}
experiments,
which have the most useful high-\pstar data at 34 and 91~\gev,
respectively.
Other data at these energies are consistent, and give the same
conclusionns.
We also generate spectra from each of the hadronization models at each
of these energies.
The pion spectra at these three energies are shown in
Fig.~\ref{fig:scal},
along with the predictions of the JETSET model.
A substantial scaling violation is visible at high $x_p$ due to the
running of the strong coupling.
Above 0.2 in $x_p$, 
the JETSET spectrum is within a few percent of our data and
also describes the data at other \ecm, 
and hence describes the scaling properties well.
The other two models also describe the scaling violation, even though
they do not describe the data well at any energy.

A similar plot for \Kpm is shown in Fig.~\ref{fig:scal}.
Here, the simulated scaling violation between 34 and 91~\gev appears
small due to the different flavor composition at the $Z^0$, 
and the 34~\gev data are of limited use.
The UCLA model is shown, as it describes our high-$x_p$ data best,
but the other models have similar scaling properties.
They predict about 15\% more scaling violation than is observed.
The data have a total experimental uncertainty of about 6\%,
but uncertainties in the flavor dependence make it difficult to
draw any conclusion.

A similar plot for \ppbar is shown in Fig.~\ref{fig:scal},
using the JETSET model with the diquark probability reduced to 0.085, 
which describes the SLD data well;
again the other models have similar scaling properties.
They predict a much larger scaling violation than is observed: 
the difference is about 80\% for 0.4$<x_p$0.7, 
and at higher $x_p$ the data from the two energies become consistent
with no scaling violation.

\begin{figure}[htb]
\centering
\includegraphics[width=0.4\hsize]{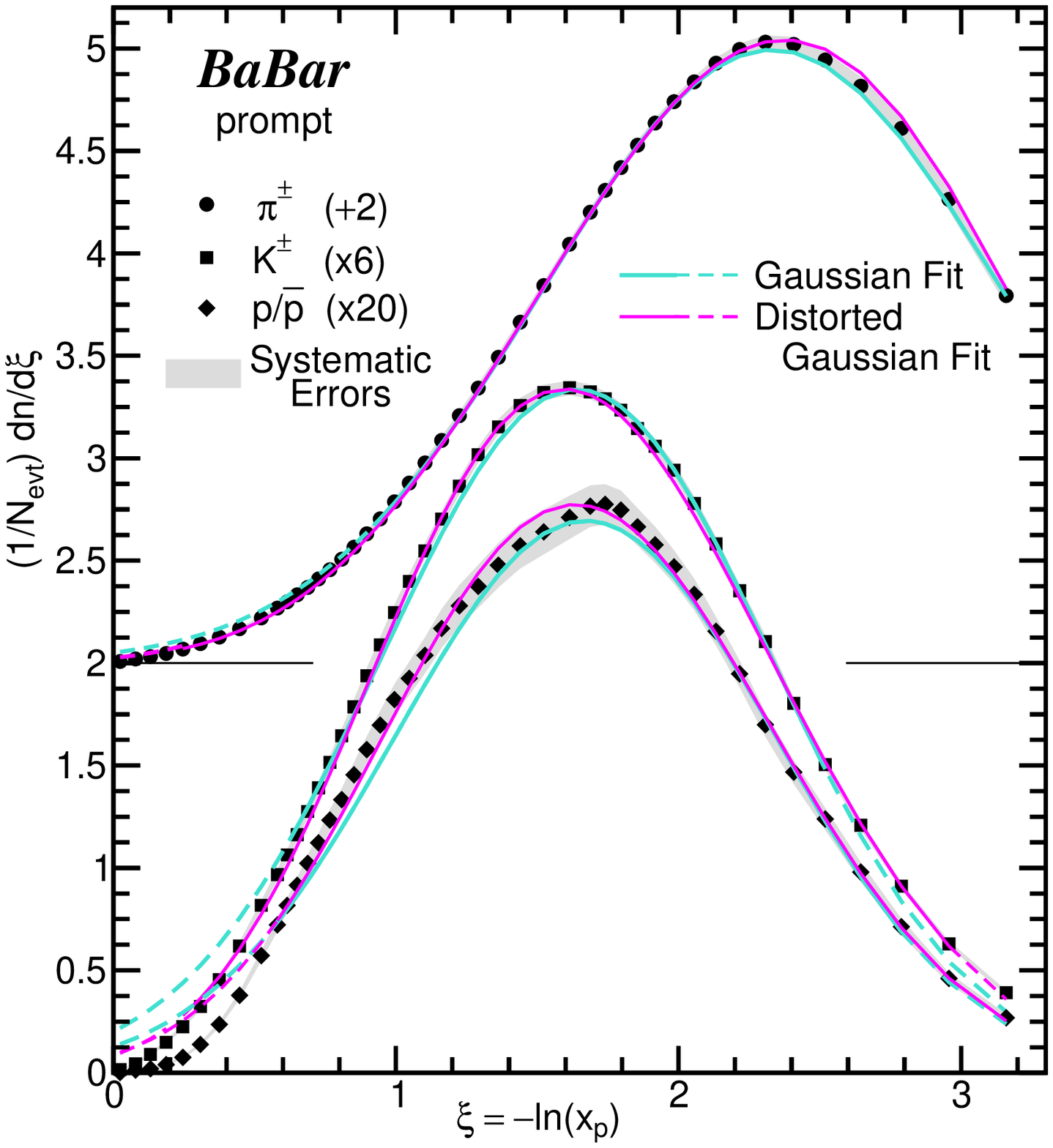}
\includegraphics[width=0.54\hsize]{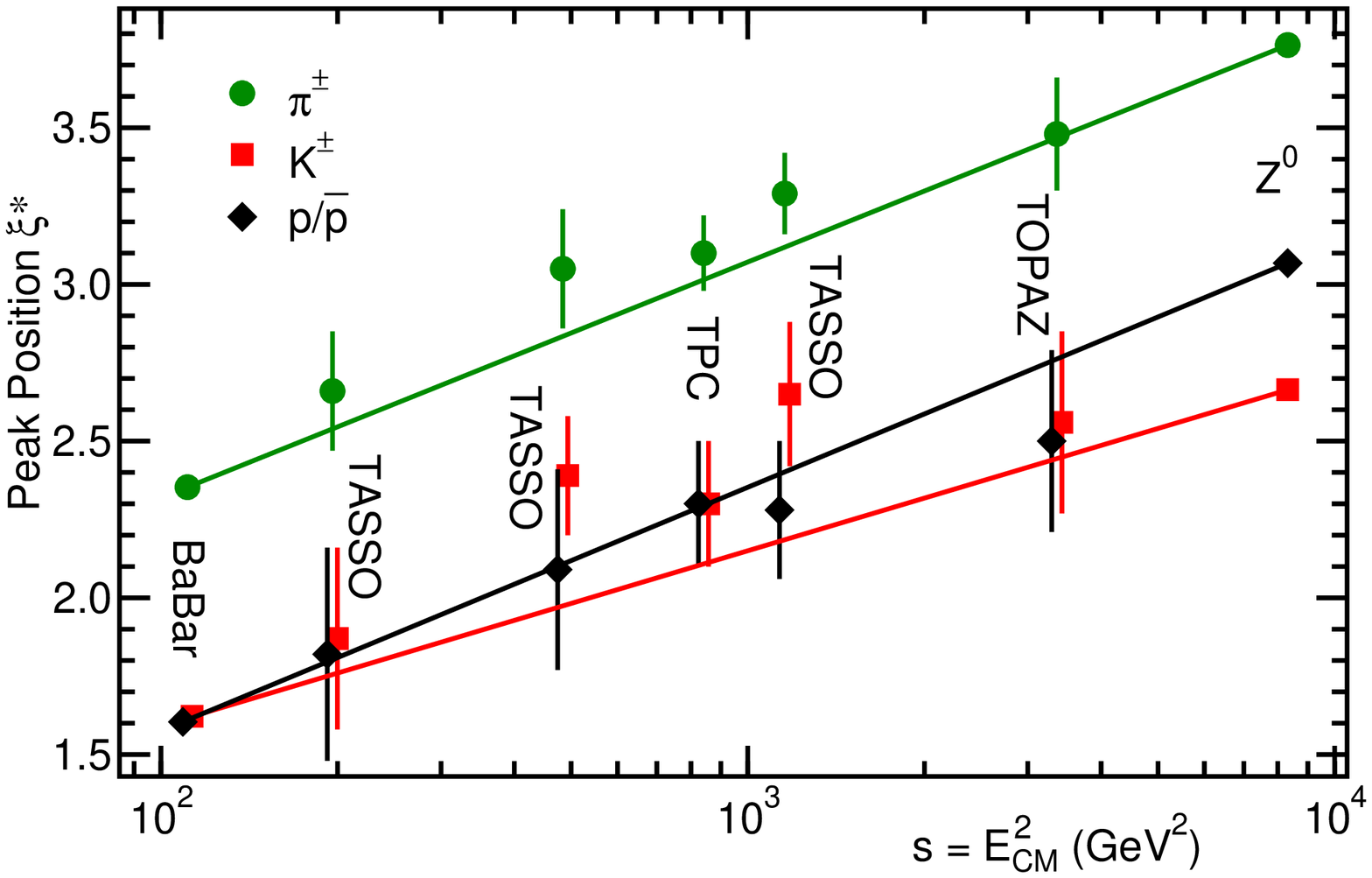}
\caption{
 Left:  
 prompt $\xi$ spectra for \pipm (circles), \Kpm (squares) and \ppbar
 (diamonds).
 The curves represent fits of Gaussian (magenta) and distorted
 Gaussian (cyan) functions over their maximal acceptable ranges;
 they are shown as solid (dashed) lines in(out)side that range.
 Right:  
 conventional peak positions $\xi^*$ vs.\ \ecm for \pipm (circles),
 \Kpm (squares) and \ppbar (diamonds) on a logarithmic horizontal
 scale. 
 The lines join our points with those from averages of results from
 the $Z^0$ experiments.
}
\label{fig:mlla}
\end{figure}

We test the predictions of QCD in the Modified Leading Logarithm
approximation, 
combined with the anzatz of local parton-hadron duality, 
using our measured spectra in the variable $\xi = -\ln x_p$.
Our $\xi$ spectra are shown in Fig.~\ref{fig:mlla};
this variable emphasizes the low-\pstar region and allows the bulk of
the spectra to be visible on a linear vertical scale.
The spectra show the slow rise from zero at $\xi = 0$ (the beam
momentum) and ``humpbacked plateau'' predicted by MLLA-QCD.
The theory predicts that the spectra should be well described by a
Gaussian function within about 1 unit of the peak position $\xi^*$,
and over a wider range by a slightly distorted Gaussian function.
We find this to be the case, 
show our fits over the widest ranges giving a good fit
(P($\chi^2$)$>$0.01) on Fig.~\ref{fig:mlla}, 
and give these ranges and $\xi^*$ values in Table~\ref{tab:summary}.

MLLA QCD also predicts that the value of $\xi^*$ should decrease
exponentially with mass for a given CM energy and 
increase logarithmically with energy for a given particle type.
Our fitted $\xi^*$ for \Kpm is lower than for \pipm (see
Table~\ref{tab:summary}), 
but that for \ppbar is not lower than for \Kpm. 
This behavior is similar to that observed at higher energies, 
where meson and baryon $\xi^*$ values each decrease with mass, 
but on different trajectories.
We compare our $\xi^*$ values with those from higher energies in
Fig.~\ref{fig:mlla}.
Our values and those from the $Z^0$ experiments are much more precise
than the rest,
and have been used to define the lines in the figure.
The other points are consistent with a logarithmic energy dependence, 
but more precise measurements are needed.
The slopes of the pion and \ppbar lines are similar, while that of the
\Kpm difffers somewhat,
perhaps due to the changing flavor composition of hadronic
events with energy.

We derive total rates per event by integrating the spectra over the
measured range and extrapolating across the unmeasured region.
The extrapolation is model dependent, and we use a combination of
hadronization models and distorted Gaussian fits to estimate
corrections and uncertainties.
This uncertainty is dominant for \pipm and substantial for \Kpm and
\ppbar. 
Results are given in Table~\ref{tab:summary}.
They are consistent with, and more precise than, previous measurements
at or near our \ecm,
and are not well predicted by any of the hadronization models.

\begin{table}[t]
\begin{center}
\begin{tabular}{ll|cccc}
 &&&& \multicolumn{2}{c}{Maximum $\xi$ range} \\
\multicolumn{2}{c|}{Particle} 
         &   Yield/Event   &      $\xi^*$    &  Gaussian  & Distorted \\
 \hline
& \pipm  &  6.07$\pm$0.16  & 2.337$\pm$0.009 & 0.92--3.27 & 0.22--3.27 \\
 Prompt
& \Kpm   & 0.972$\pm$0.020 & 1.622$\pm$0.006 & 0.63--2.58 & 0.34--3.05 \\
& \ppbar & 0.185$\pm$0.006 & 1.647$\pm$0.019 & 0.56--3.27 & 0.48--3.27 \\
 \hline
& \pipm  & 6.87 $\pm$0.19  & 2.353$\pm$0.009 & 0.87--3.27 & 0.67--3.27 \\
 Conventional 
& \Kpm   & 0.972$\pm$0.020 & 1.622$\pm$0.006 & 0.63--2.58 & 0.34--3.05 \\
& \ppbar & 0.265$\pm$0.008 & 1.604$\pm$0.013 & 0.71--2.58 & 0.48--3.27 \\
\hline
\end{tabular}
\caption{The average number of each particle type produced per
  hadronic \epem event at 10.54~\gev.
  The position of the peak of each $\xi$ distribution, along with the
  widest range over which a Gaussian or distorted Gaussian fit is
  able to describe the spectrum.}
\label{tab:summary}
\end{center}
\end{table}

\section{Summary}

In summary,
we have measured the differential and total 
inclusive production rates for \pipm, \Kpm and
\ppbar in hadronic \epem annihilations at 10.54~\gev,
both excluding (prompt) and including (conventional) the decay
products of \KS and weakly decaying strange baryons.
The measurements cover the range from 0.2~\gevc to the kinematic
limit,
and improve upon the precision and coverage of previous measurements
at or near this \ecm.

These results can be used to test and tune hadronization models.
None of the models tested is able to describe the shape or amplitude
of the data with its
default parameter settings.
With higher energy data, the scaling properties of the hadronization
process can be studied.
The models are able to describe the high-$x_p$ scaling of \pipm from 10.54 to
91.2~\gev well, but fail to describe that for \ppbar.
The shapes of the spectra at low \pstar (high $\xi$) are well
described by MLLA QCD, as well as the dependence of the peak position
$\xi^*$ on \ecm and hadron mass, except that that for \ppbar is not
lower than that for \Kpm.


\end{document}

%% file: econfmacros.tex
\def\beq{\begin{equation}}
\def\eeq#1{\label{#1}\end{equation}}
\def\eeqn{\end{equation}}

\def\beqa{\begin{eqnarray}}
\def\eeqa#1{\label{#1}\end{eqnarray}}
\def\eeqan{\end{eqnarray}}

\let\bar=\overbar

\def\Dslash{\not{\hbox{\kern-4pt $D$}}}
\def\dslash{\not{\hbox{\kern-2pt $\del$}}}

\def\BR{\mbox{\rm BR}}

\def\msb{{\bar{\ssstyle M \kern -1pt S}}}




%% file: DPF13-muller.bbl
\begin{thebibliography}{99}


\bibitem{babarpikp}
J.P.~Lees, et al. (\babar\ Collaboration), 
Phys.\ Rev.\ D {\bf 88}, 032011 (2013).

\bibitem{arguspikp}
H.~Albrecht et al. (ARGUS Collaboration),
Z.\ Phys.\ C {\bf 44}, 547 (1989).

\bibitem{tassopikp}
W.~Braunschweig et al. (TASSO Collaboration),
Z.\ Phys.\ C {\bf 42}, 189 (1989).

\bibitem{sldpikp}
K.~Abe et al. (SLD Collaboration),
Phys.\ Rev.\ D {\bf 69}, 072003 (2004).


\end{thebibliography}
